\documentclass[prl,twocolumn,showpacs,amsmath,amssymb,superscriptaddress]{revtex4-1}
\pdfoutput=1
\usepackage[usenames,dvipsnames]{color}
\usepackage{longtable}
\usepackage{epsfig}
\usepackage{amssymb}
\usepackage{amsmath}
\usepackage{graphicx}
\usepackage{verbatim}
\usepackage{xspace}

\def\araa{ARA\&A}%
\def\apj{ApJ}%
\def\apjl{ApJ}%
\def\apss{Ap\&SS}%
\def\aap{A\&A}%
\def\jcap{JCAP}%
\def\mnras{MNRAS}%
\def\nat{Nature}%
\definecolor{darkblue}{rgb}{0,0,0.5}
\definecolor{darkgreen}{rgb}{0.1,0,0.3}
\definecolor{darkred}{rgb}{0.6,0,0}

\newcommand{\ba}{\begin{eqnarray}}
\newcommand{\ea}{\end{eqnarray}}
\newcommand{\be}{\begin{equation}}
\newcommand{\ee}{\end{equation}}

\newcommand{\gn}{G}

\newcommand{\dgt}{{\dot{G}/G}_{\rm today}}
\newcommand{\sz}{S_{pp}(0)}
\newcommand{\yr}{{\rm yr}}

\newcommand{\CODATA}{\texttt{CODATA} }
\newcommand{\BIS}{\texttt{BiSON} }

\newcommand{\OPAL}{\texttt{OPAL} }
\begin{document}
\title{A new helioseismic constraint on a cosmic-time variation of G}

\author{Alfio Bonanno}
 \affiliation{INAF, Osservatorio Astrofisico di Catania, via S. Sofia, 78, 95123 Catania, Italy}

\author{Hans-Erich  Fr\"ohlich}
 \affiliation{Leibniz Institute for Astrophysics Potsdam (AIP), An der Sternwarte 16, 14482 Potsdam, Germany}

\begin{abstract}
Helioseismology can provide strong constraints on the evolution of Newton's constant over cosmic time. 
We make use of the best possible estimate of 8640 days of low-$\ell$ \BIS data, corrected for the solar
cycle variation, to obtain a new constraint on an evolving gravitational constant. 
In particular, by means of a Bayesian analysis we conclude that $\dgt=(1.25\pm 0.30) \times 10^{-13} \; \yr^{-1}$. 
Our result, a 4-$\sigma$ effect, is more than one order of magnitude stronger than previous constraints obtained with helioseismology.
We also take into account  possible systematic effects by considering the theoretical uncertainties on the efficiency 
of the proton-proton $(pp)$ fusion cross-section. We show that models with variable $G$ significantly outclass models with no secular variation
of $G$, viz by a Bayes factor exceeding 30. 
\end{abstract}

\maketitle

\section{Introduction}
The idea that the Sun can be considered a laboratory for fundamental physics traces back 
to the early developments in nuclear physics by contributing to the understanding 
of the basic nuclear processes involved in stellar nucleosynthesis.
In recent times, accurate measurements of acoustic $p$-mode spectrum combined with inversion 
techniques have further stressed this role \cite{sarbani16}.
Important examples are the investigation of the equation of state \citep{basu99}, 
the discovery of neutrino flavour oscillations \citep{fukuda,sno}, 
the properties of Dark Matter \citep{lopes02,lopes04,lopes12,aldo15}, 
the constraints on axions emission \citep{helmut99,aldo15b}
the properties of the screening of nuclear reaction rates \citep{fiorentini01,weiss01} 
and constraints on physical constants \citep{joergen05}. 

A fundamental problem that can be tackled by means of  helioseismology is the possibility of limiting secular variations of
$\gn$, a possibility argued long ago by 
Dirac \citep{dirac38} and Milne \citep{milne37}.  
This initial intuition has been further elaborated in \cite{brans61,bergmann} and is nowadays an important ingredient of  
various scalar-tensor theories \citep{maeda},  quantum-gravity inspired models of modified gravity \citep{bo04,smolin2015}, 
and string theory low-energy models \cite{gaspe}.

In this context a widely used approach to promote the gravitational constant to a dynamical variable is to extend
the general relativistic framework in which gravity is mediated by a massless spin-2 graviton, 
to include a spin-0 scalar field which couples universally to matter fields. 
As the universality of free-fall is maintained 
theories that predict that the locally measured gravitational constant vary with time
often violate the equivalence principle in its strong form. For this reason 
empirical constraints on $\dgt$, where the dot indicates a derivative with respect to the cosmic time $t$, 
have been obtained in several contexts \citep{uzan11,peebles16}. 
Current limits on $\dgt$ span from $\dgt=(4\pm 9) \times 10^{-13} \yr^{-1}$  
obtained from the Lunar Laser Ranging (LLR) experiment \citep{llr04}, to 
$-3\times 10^{-13}< \dgt < 4 \times 10^{-13} \yr^{-1}$ 
from BBN  \cite{bbn04}, or $\dgt\sim 10^{-12} \yr^{-1}$ from white dwarfs \citep{dwarfs}. 

Helioseismology is able to provide independent constraints on possible time evolution 
of the gravitational constant $G$ over cosmic time
because the stellar luminosity $L$ varies as  $\sim G^{\,7}$ \cite{scilla96}. 
For example, a monotonically increasing  Newton's constant  must be compensated for a systematic decrease of 
core temperature and a corresponding change in the hydrogen abundance in order to match
$L_\odot$, the solar radius $R_\odot$ and the metal to hydrogen abundance ratio $(Z/X)_\odot$. 
In \citep{gue98} a direct comparison of low-degree $p$-modes to GONG data has allowed us 
to obtain $\dgt \leq 1.6 \times 10^{-12}\yr^{-1}$,  
assuming a power-law of the type  $G(t)\propto t^{-\alpha}$. 
In this paper we shall present a new limit  on $\dgt$ based on a bayesian approach which makes use
of the definitive ``best possible estimate'' of  8640 days of low-$\ell$ frequency \BIS data, corrected for the
solar cycle modulation \citep{broom}.

\begin{figure}[tb]
\includegraphics[width=.5\textwidth]{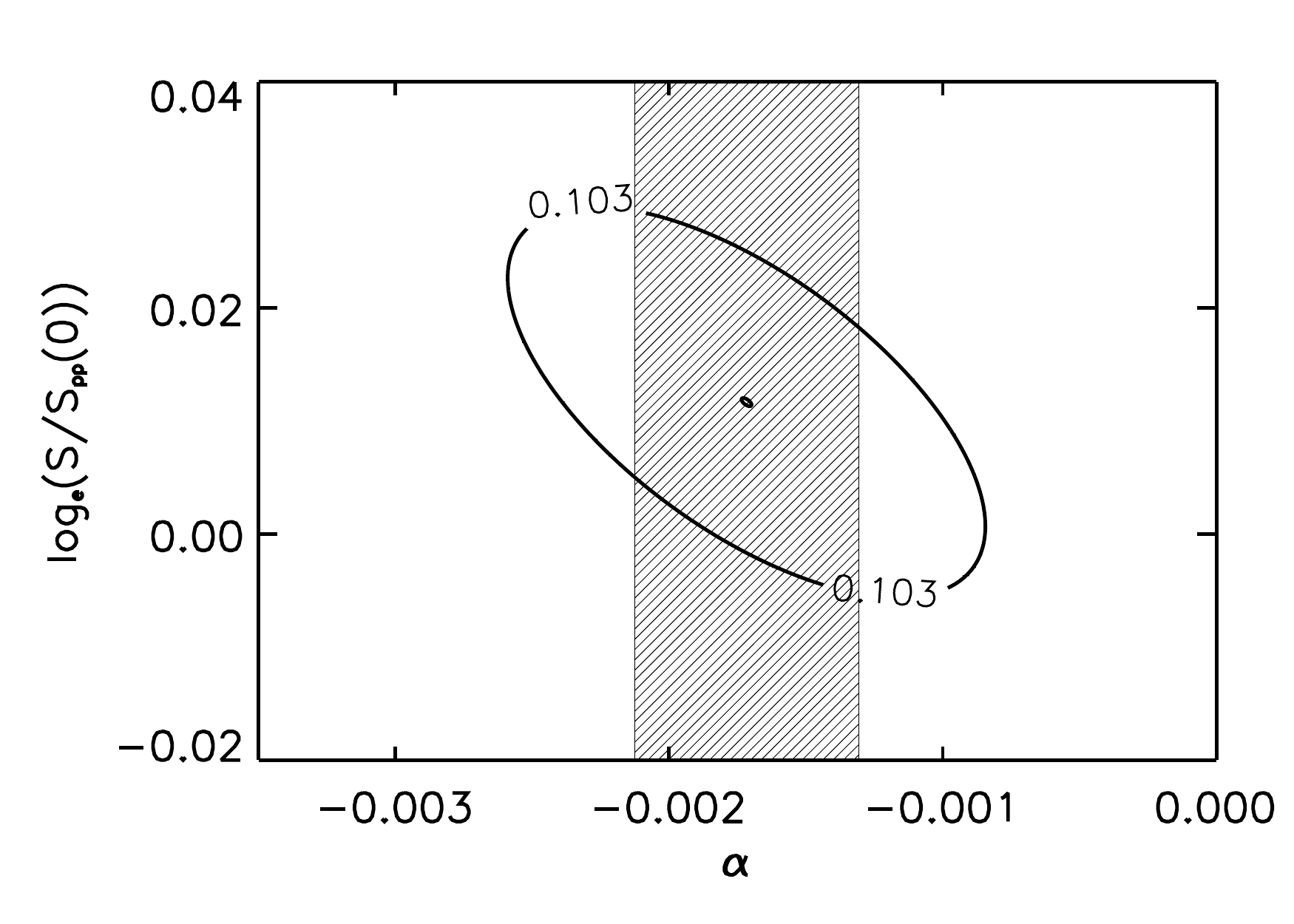}
\caption{2-dimensional posterior probability distribution. The closed contour with a probability 
density of 10.3  per cent of the peak density comprises 90 per cent of  the total probability. 
The hatched strip marks the 68.3-per-cent ($\pm$ 1-$\sigma$) interval of $\alpha$'s marginal distribution. }
\label{fig1}
\end{figure}

\section{Solar models and model uncertainties}
In this context it is important to reduce as much as possible any source of systematic uncertainties in the input physics 
of the calibrated solar models in order to obtain a significant constraint on $\dgt$. 
From this point of view the main problem
is clearly our ignorance of the efficiency of the 
proton-proton ($pp$) fusion cross-section for which only theoretical estimates are 
available.  An uncertainty of $\pm 3\%$ on the value of $\sz$, the astrophysical $S$-factor at zero energy,
is quoted in \citep{adelberger}, in particular. Therefore both 
$\dgt$ as well as $\sz$, have been estimated from the data in a Bayesian manner. 

Our solar models are built using the Catania version of the GARSTEC code
\citep{bosch02,garstec}, a fully-implicit 1D code including heavy-elements diffusion and updated input physics. 
We prescribed the time evolution of the gravitational constant as a power-law \citep{uzan,gue98}
\be\label{gv}
G(t)=G_0 \left (\frac{t_0}{t} \right)^\alpha 
\ee
where $G_0$ is the cosmologically recent value of Newton's constant according to 2010 \CODATA
so that $G_0=6.67384 \times 10^{-8}$ ${\rm cm^3 g^{-1}s^{-2}}$
and $t_0=13.7$ Gyr is a reference age of the Universe according to most of $\Lambda$CDM estimates.
As  $ G_0 M_\odot \equiv 1.32712440 \times 10^{26}$ ${\rm cm^3 s^{-2}}$
\citep{cox} is fixed, $M_\odot = 1.98855 \times 10^{33} \rm g$ is assumed. 
Irwin's equation of state \citep{cassisi} with
\OPAL opacities for high temperatures \citep{iglesias}
and Ferguson's opacities for low temperatures \citep{ferguson} are employed and 
the nuclear reaction rates are taken from the compilation in \cite{adelberger}. 

Our starting models are chemically homogeneous PMS models with $\log L/L_\odot = 0.21$ and
$\log T_e = 3.638 $ K, thus close to the birth line of a $1 M_\odot$ object.
Initial Helium fraction, $(Z/X)$ and mixing-length parameter are adjusted to match 
the solar radius $R_\odot = 6.95613 \times 10^{10} \rm cm$
(based on an average of the two values and quoted error bar in Table
3 of \cite{haber}), the solar luminosity $L_\odot = 3.846 \times 10^{33}~\rm erg~s^{-1}$ \citep{cox}
and the chemical composition of \cite{gn93} with 
$(Z/X)_\odot=0.0245$ at the surface. 
We also employed the new accurate meteoritic estimate of the solar age of \cite{conn}, $t_\odot = 4.567$ Gyr, 
a value consistent with the helioseismic solar age \citep{bo15}.
We further noticed that models with the   so-called ``new abundances''
for which  $(Z/X)_\odot=0.0178$ \citep{asplund09}
would lead to much smaller Bayes factors and we decided not to discuss these models
in this work.

In order to define a proper seismic diagnostic we adopted a widely used approach:
if $\nu_{n,l}$  is the frequency of the mode of radial order $n$ and angular degree $\ell$,
the frequency separation ratios
\be\label{sep}
r_{l,l+2}(n) = \frac{\nu_{n,l}-\nu_{n-1,l+2}}{\nu_{n,l+1}-\nu_{n-1,l+1}}
\ee
can be shown to be localized near the core and weakly dependent on the complex physics of 
the outer layers \citep{rox03,oti05}.  In particular  in  the limit $n\gg 1$ 
\begin{align}
r_{\ell,\ell+2}(n) \approx -(4\ell+6) \frac{1}{4\pi^2 \nu_{n,\ell}}\int_0^{R_\odot}{\frac{dc_s}{dR}\frac{dR}{R}}
\end{align}
so that a change  in temperature 
($T$) and  mean molecular weight ($\bar\mu$) directly impacts on the $r_{\ell,\ell+2}(n)$ terms as
$\delta c_s/c_s \approx \frac{1}{2} \delta T/T - \frac{1}{2} \delta \bar\mu/\bar\mu$. 

\section{Bayesian approach}

We consider the following two-dimensional parameter space: $-0.1 \le \alpha \le 0.1$ and 
$0.97 \le S/\sz \le 1.03$. 
The proposed $\alpha$ range generously covers all previous $\dgt$ limits 
obtained by independent methods \citep{uzan}. 
Moreover, the $S$ interval 0.97--1.03 allows for an up to $\pm 3\%$ deviation from the recommended value
$\sz=(4.01\pm 0.04) \times 10^{-22}$ keV b in \cite{adelberger}.


Central to the Bayesian hypothesis testing is the likelihood. 
In the following, a Gaussian has been assumed, 
\begin{equation}
{\cal L}(\alpha, S) =\prod_{i=1}^{N=17}\frac{1}{\sqrt{2\pi}\sigma_i}\exp\left(-\frac{(d_i-m_i(\alpha,S))^2}{2\sigma_i^2}\right)\,,
\label{l1}
\end{equation}
where $d_i=r_{02}(n)$ are the observed data ($n=i+8, i=1\dots{}N, N=17$), 
$m_i$ the theoretical model values, and $\sigma_i$  the errors 
(see also \citep{bo15} for an application of this likelihood 
to the helioseismic determination of the solar age). 
All 17 contributions enter the likelihood with the same weight. 

The posterior probability distribution is the likelihood \eqref{l1} weighted with a prior distribution. 
Obviously, this prior distribution should be a flat one compared to $\alpha$. Concerning $S$ we decided 
to take a conservative point of view, i.\,e. that nothing is known about $\sz$. In that case 
we are on the safe side and the only eligible prior distribution is a flat one over the logarithm, $\log(S)$. 

In the end two hypotheses have to been compared: H$_1$ = H$_1(-0.1\le\alpha\le 0.1,0.97\le S/\sz \le 1.03)$ vs. 
our zero hypothesis H$_0(\alpha=0,0.97 \le S/\sz \le 1.03)$.  
\section{Results}
The posterior probability distribution is indistinguishable from a two-dimensional Gaussian (Fig.~\ref{fig1}). 
The reason is that the theoretical models $m_i(\alpha,\log(S))$ are linearly  
dependent on both $\alpha$ and  $\log(S)$ (cf. \citep{kendall2B}) as we checked in all our models. 
From $\alpha$'s marginal distribution one reads its mean value and standard deviation: 
$\langle \alpha \rangle = -0.0017\pm 0.0004$. Formally, this is a 4-$\sigma$ effect.
With $t_0 = 13.7$\,Gyr this translates to $\dgt = (1.25 \pm 0.30) \times 10^{-13}\,\yr^{-1}$. 
As a by-product one gets $\langle \log(S/\sz) \rangle = 0.011\pm 0.008$ and a correlation coefficient of -0.62.
An enhanced $S$ goes with a reduced $\alpha$. However, the indicated slight enhancement of 
Adelberger et al. \citep{adelberger} $pp$ cross-section by 1\% proves insignificant. 
Our result is  one order of magnitude stronger than the limit obtained in \cite{gue98} and comparable in 
precision to those obtained with LLR \cite{llr04}  or BBN \cite{bbn04}.

Integrating the posterior over the whole parameter space or subsections of it, respectively, 
one gets the required evidences. The evidence in favour of a hypothesis is 
the prior-weighted mean of the likelihood over parameter space. 
The ratio of the evidences, E(H$_1$)/E(H$_0$), the so-called Bayes factor amounts to 34.0. 
(If one trusts the relative error in the recommended $\sz$ and applies the appropriate 
Gaussian prior, this Bayes factor would increase to 51.1.)
Despite one parameter more, the $\alpha\ne 0$ hypothesis significantly outclasses 
the zero hypothesis, i.\,e. no secular variation of $G$ -- provided 
the $S$ factor is the sole and decisive unknown. 

\bigskip
\textbf{\textit{Acknowledgements.---}}
We acknowledge L. Santagati for careful reading of the manuscript.

\end{document}